# Monopole-like orbital-momentum locking and the induced orbital transport in topological chiral semimetals

Qun Yang[a,b*], Jiewen Xiao[b], Iñigo Robredo[a,c], Maia G. Vergniory[a,c], Binghai Yan[b*], and Claudia Felser[a*]

[a]Max Planck Institute for Chemical Physics of Solids, 01187 Dresden, Germany

[b]Department of Condensed Matter Physics, Weizmann Institute of Science, Rehovot 7610001, Israel

[c]Donostia International Physics Center, 20018 Donostia-San Sebastian, Spain

*Corresponding authors: qun.yang@cpfs.mpg.de, binghai.yan@weizmann.ac.il, claudia.felser@cpfs.mpg.de

**The interplay between chirality and topology nurtures many exotic electronic properties. For instance, topological chiral semimetals display multifold chiral fermions that manifest nontrivial topological charge and spin texture. They are an ideal playground for exploring chirality-driven exotic physical phenomena. In this work, we reveal a monopole-like orbital-momentum locking texture on the three-dimensional Fermi surfaces of topological chiral semimetals with B20 structures (e.g., RhSi and PdGa). This orbital texture enables a large orbital Hall effect (OHE) and a giant orbital magnetoelectric (OME) effect in the presence of current flow. Different enantiomers exhibit the same OHE which can be converted to the spin Hall effect by spin-orbit coupling in materials. In contrast, the OME effect is chirality-dependent and much larger than its spin counterpart. Our work reveals the crucial role of orbital texture for understanding OHE and OME effects in topological chiral**

semimetals and paves the path for applications in orbitronics, spintronics, and enantiomer recognition.

**Keywords:** chirality, topology, orbital-momentum locking, orbital Hall effect, magnetoelectric effect, enantiomer recognition

**Introduction**

Chirality refers to a spatial asymmetric feature enabling objects to appear in two non-superimposable mirror-image forms (enantiomers). Opposite enantiomers may exhibit different and even completely opposite physical/chemical properties and behave differently in response to external stimuli. This has brought significant consequences and applications in chemistry and biology (1), for example in enantioselective catalysis and drug design. In condensed matter physics, due to the rich interplay between chiral symmetry, relativistic effects, and electronic transport, the chiral solid crystals constitute an ideal playground for exploring exotic physical phenomena. A variety of chirality-driven unconventional electronic transport, such as electrical magnetochiral-anisotropy (2), chirality-induced spin selectivity (3–7), and unidirectional magnetoresistance (8) have been found in different systems. It is of fundamental interest to broaden the spectrum of chirality-driven physical properties and related phenomena that can lead to practical applications.

Recently, chirality was found to induce topological electronic property with orbital-momentum locking in DNA-like molecules (9) and results in chirality-induced spin selectivity (5) and anomalous circularly polarized light emission (10). Compared to molecules, chiral solid crystals may exhibit more intriguing orbital texture that brings exotic physical/chemical phenomena, yet to be explored. The topological chiral semimetals in the space group P $2_1$ 3, such as PdGa, PtAl, and CoSi *et al* (11–15), which are characterized by multifold band crossings with large Chern numbers in the bulk state and unique Fermi arcs at surfaces, are of particular interest. These crystals display promising applications in chiral catalysis (15–18). Exploring the orbital effect is also helpful to understand the relation between the band structure topology and enantioselective processes, such as enantiomer recognition or chiral catalysis.

In this work, we study the orbital texture in the band structure of topological chiral semimetals. The results indicate that for the opposite enantiomers of topological chiral semimetals, the sign reverses in orbital polarization. Near the multifold band crossing points, the orbital angular momentum (OAM) texture exhibits monopole-like characteristics, similar to the Berry curvature distribution of Weyl points in momentum space. We demonstrate that such OAM texture leads to a large orbital Hall effect (OHE) and orbital magnetoelectric (OME) effect, which is insensitive to the spin-orbital coupling (SOC) strength of the materials. Among all five topological chiral semimetals that we studied, RhSi and PdGa are the most promising candidates for detecting the large orbital Hall conductivity (OHC) and current-induced orbital magnetization in the experiment. Furthermore, we point out that OHE is enantiomer-independent while the sign of the OME effect manifests the chirality. Our study reveals the exotic electronic orbital texture induced by chirality and indicates a way to utilize the orbital transport for applications in orbitronics, spintronics, and enantiomers recognition.

**Results and Discussion**

The widely studied topological multifold semimetals, such as CoSi, RhSi, PdGa, PtAl and PtGa crystallize in the B20 (FeSi-type) crystal structure with structural chirality (11–14). They belong to the Sohncke non-symmorphic space group (SG) P$2_1$3 (No. 198) generated by twofold screw rotations $2_{1x} = \{C_{2x} | 0.5, 0.5, 0\}$, $2_{1y} = \{C_{2x} | 0, 0.5, 0.5\}$ and diagonal threefold rotations $C_{3,111} = \{C_{3,111} | 0.0, 0.0, 0\}$. The combination of these symmetry operations give rise to three two-fold screw rotation axes along the axes of the Cartesian coordinate system and four three-fold rotation axes along the cube's main diagonals. The material is non-magnetic; therefore, the time-reversal symmetry (TRS) and its combination with other crystal symmetries also belong to the symmetry group. Figure 1a depicts the crystal structures of two PdGa enantiomers (enantiomers A and B) with winding of the helices related by an inversion operation, where structural chirality determines electronic properties. Figure 1b and Figure 1c show symmetry-protected multifold band crossings appearing at the Γ and R points in the bulk, which carry the Chern numbers of -4 and 4, respectively. The sign is reversed in the opposite enantiomer.

Herein, we present the detailed symmetry analyses of the OAM in topological chiral semimetals. Despite the atomic orbital being quenched in solids, the OAM can be estimated based on Bloch wave functions. The general $L_n^\gamma(\mathbf{k})$ component of the OAM for band n at point $\mathbf{k}$ is given by (19–21):

$$L_n^\gamma(\mathbf{k}) = 2\epsilon_{\alpha\beta\gamma}\hbar m_e \mathrm{Im} \sum_{m \neq n} \frac{\langle n(\mathbf{k})|\hat{v}_\alpha|m(\mathbf{k})\rangle\langle m(\mathbf{k})|\hat{v}_\beta|n(\mathbf{k})\rangle}{E_n(\mathbf{k}) - E_m(\mathbf{k}) + i\eta}, \tag{1}$$

where $\alpha, \beta, \gamma = x, y, z$, $\epsilon_{\alpha\beta\gamma}$ is the Levi-Civita symbol. $E_n(\mathbf{k})$ is the eigenvalue for the $n_{th}$ eigenstate of $|n(\mathbf{k})\rangle$ at the momentum $\mathbf{k}$. $\hat{v}_{\alpha(\beta)}$ is the $\alpha(\beta)$ component of the band velocity operator with $\hat{v}_{\alpha(\beta)} = \frac{1}{\hbar}\frac{\partial \hat{H}(\mathbf{k})}{\partial k_{\alpha(\beta)}}$, $\hat{H}$ is the Hamiltonian operator, and $\eta = 0.1$ meV represents a very small broadening for numerical reasons.

OAM is a pseudovector that transforms like spin under symmetry operations. The key OAM relationships for the materials in SG $P2_13$, are listed in Table S1, which define the constraints on the OAM in the k-space. Particularly, the OAM of the opposite enantiomers is related by the inversion operation as follows: $\mathbf{L}^A(\mathbf{k}) \to \mathbf{L}^B(-\mathbf{k})$, $\mathbf{L}^B(-\mathbf{k}) = -\mathbf{L}^B(\mathbf{k})$, where the latter equality is enforced by TRS. This relation shows that the OAM changes sign on opposite enantiomers. As a consequence of TRS, the OAM changes sign on reversing the momentum. Furthermore, at any time-reversal invariant momenta, the OAM must vanish since the OAM changes sign while the momentum is left invariant. Away from the high-symmetry points, the symmetry relations for various OAM components along the high-symmetry lines or planes can be further discussed. For instance, the $\Gamma - X$ high-symmetry line is left invariant by the $2_{1x}$ axis. Therefore, we have $L_x(k_x, 0,0) \neq 0$, $L_y(k_x, 0,0) = 0$, and $L_z(k_x, 0,0) = 0$. This relation can be translated to the other X-point-directed paths ((0, $k_y$, 0) and (0, 0, $k_z$)) using permutation of the indices. Another instance is of the Γ-R high-symmetry line which remains invariant under the $C_3$ operation: $L_x(k, k, k) = L_y(k, k, k) = L_z(k, k, k)$. More detailed OAM symmetry properties for the materials in SG P$2_1$3 can be deduced directly from Table S1.

We then performed the density-functional theory (DFT) calculations (22, 23) (see METHODS section for calculation details) for realistic bulk compounds. Since the symmetry of Wannier functions is essential for the real material OAM calculations, we projected the ab initio DFT Bloch wavefunction into highly symmetric atomic-orbital-like Wannier functions and generated the corresponding tight-binding (TB) model Hamiltonian that fully respects the symmetry of corresponding materials. Using the obtained TB Hamiltonian, the OAM can be numerically computed based on the Eq. (1). Figure 2a shows the OAM-resolved band structure along the $R - \Gamma - R$ high symmetry line for the two PdGa enantiomers. The OAM-resolved band structures for other topological chiral semimetals including CoSi, RhSi, AlPt and PtGa, along different momentum directions [001], [111], and [110], are shown in Figure S1. As expected, ab initio calculations are in good agreement with the expected OAM symmetry properties: the orbital polarization reverses signs at opposite momentum and the OAM texture changes sign for the opposite enantiomers. Furthermore, owing to the non-centrosymmetric crystal structure of the material, SOC splits the energy bands. The spin-split bands carry the opposite spin texture and approximately the same OAM, which can be observed from the plot of the OAM vector fields for spin-split Fermi surface (FS) pairs as shown in Figure S2.

The topological chiral semimetal PdGa displays a nontrivial momentum dependence of the Berry curvature at multifold fermions $\Gamma$ and $R$ (Figure 1b). Since the OAM has an intimate connection to the Berry curvature (24), it is of fundamental interest to study the OAM texture distributions near $\Gamma$ and R. Figure 2b shows the Fermi pockets that are around 30 meV above the chiral fermions $\Gamma$ and R, respectively, which are formed by the band $\gamma_2$ in Figure 1c. Moreover, Figure S3 shows the FS pockets near the chiral fermions. The OAM texture was found to be similar to the Berry curvature monopole. In enantiomer A, the OAM texture exhibits a monopole-like feature at $\Gamma$ (source) and $R$ (drain). As expected, this OAM texture reverses in enantiomer B.

The nontrivial OAM texture in topological chiral semimetals further induces orbital transport phenomena. Here we focus on OHE and magnetoelectric (ME) effect. OHE (25) is a phenomenon of the generation of transverse OAM current in response to an applied electric field, which is an orbital analog to the transverse spin angular momentum (SAM) current in the spin Hall effect (SHE) (26, 27). In OHE, the applied electric field along β-direction ($E_\beta$) and the induced orbital current along α-direction with orbital polarization along γ-direction ($J_\alpha^\gamma$) are related by the OHC tensor ($\sigma_{\alpha\beta}^\gamma$) as $J_\alpha^\gamma = \sigma_{\alpha\beta}^\gamma E_\beta$. With the material-specific TB Hamiltonian, the $\sigma_{\alpha\beta}^\gamma$ can be calculated by the Kubo formula (25):

$$\sigma_{\alpha\beta}^{\gamma} = e\hbar \int_{BZ} \frac{d\mathbf{k}}{(2\pi)^3} \sum_n f_{n\mathbf{k}} \Omega_{n,\alpha\beta}^{\hat{L}_\gamma}(\mathbf{k}),$$

$$\Omega_{n,\alpha\beta}^{\hat{L}_\gamma}(\mathbf{k}) = -2\text{Im} \sum_{m \neq n} \frac{\langle n(\mathbf{k})|\hat{j}_\alpha^{\hat{L}_\gamma}|m(\mathbf{k})\rangle \langle m(\mathbf{k})|\hat{v}_\beta|n(\mathbf{k})\rangle}{(E_n(\mathbf{k}) - E_m(\mathbf{k}))^2}. \quad (2)$$

Here, $\hat{j}_\alpha^{\hat{L}_\gamma} = \frac{1}{2}\{\hat{v}_\alpha, \hat{L}_\gamma\}$ is the conventional orbital current operator, $\Omega_{n,\alpha\beta}^{\hat{L}_\gamma}(\mathbf{k})$ is referred to as the orbital Berry curvature, $\hat{L}_\gamma$ is the orbital operator and $f_{n\mathbf{k}}$ is the Fermi-Dirac distribution function. For the integral in Eq. (2), the **k**-space integration was performed on a uniform 240×240×240 k-grid.

To illustrate the general properties of the OHE for all chiral structures, we present the OHC tensor for 11 chiral point groups using the TENSOR program from the Bilbao Crystallographic Server (28), as shown in Table S2. Specially, topological chiral semimetals belong to the T point group. The existing symmetries force many tensor elements of $\sigma_{\alpha\beta}^{\gamma}$ to be zero and relate them to each other as $\sigma_{xy}^z = \sigma_{yz}^x = \sigma_{zx}^y$ and $\sigma_{yx}^z = \sigma_{zy}^x = \sigma_{xz}^y$, leaving only two groups of non-zero elements. Figure 3 shows the calculated OHC as well as SHC for five selected topological chiral semimetals, in which the Fermi energy lies at the charge neutral point. We found that the magnitude of OHC in these compounds is in general gigantic and insensitive to SOC, reaching approximately $\sim 3000 \, (\hbar/e)(\Omega \, \text{cm})^{-1}$. In contrast, the SHC shows a clear correlation with the SOC strength for different compounds, whose value is of one order of magnitude smaller than OHC. This reflects that the OHE is more robust and can be converted into SHE when the SOC is present. Furthermore, we point out that different enantiomers exhibit the same OHE/SHE since the inversion operation does not change the sign of the OHC/SHC tensor.

To elucidate the origin of the OHC change in different compounds, we show the energy-dependent OHC $\sigma_{xy}^z$ in Figure 4a and $\boldsymbol{k}$-resolved orbital Berry curvature $\Omega_{xy}^z$ of the $n^{th}$ band in Figure 4b, where red (blue) denotes a positive (negative) contribution. As shown clearly, for CoSi and RhSi, at the charge neutral point, the hole pockets near Γ and electron pockets near R make a dominant contribution. RhSi exhibits larger $\Omega_{xy}^z$ near Γ than CoSi and therefore shows stronger $\sigma_{xy}^z$. For PtAl, PtGa, and PdGa, the hole bands at M shift above the Fermi energy compared to RhSi/CoSi, attributing to different natures between Pt/Pd-d and Rh/Co-d orbitals and consequently their electron bands at Γ shift down to balance the charge neutrality. Because electron bands at Γ contribute negative $\Omega_{xy}^z$, PtAl, PtGa, and PdGa exhibit smaller $\sigma_{xy}^z$ than RhSi.

Our results reveal a large orbital Hall response in topological chiral semimetals, which is insensitive to the SOC strength, making these materials excellent candidates for detecting the OHE. It was theoretically demonstrated that the orbital current in OHE can be converted to the spin current via the SOC from the contact (29). Furthermore, the OHE can also generate large nonreciprocal magnetoresistance when employing magnetic contact. These strategies pave the way to probe the OHE in experiments. Noteworthy, the recent experiment confirmed this prediction by measuring a large effective spin Hall angle in Cu and Al through OHE to SHE conversion by the interfacial SOC (30). This will inspire experimental detection of OHE in chiral semimetals proposed in this work.

In addition to the OHE, the intrinsic OAM texture in topological chiral semimetal is also crucial for the OME effect. Generally, the ME effect requires an applied electric field in the *j*-direction ($E_j$) to induce magnetization in the *i*-direction ($M_i$), which are related *via* response coefficients $a_{ij}$ by the response equations $M_i = \sum_{i,j} a_{ij} E_j$. $a_{ij}$ is the ME susceptibility, which can be derived from the standard linear response theory as follows (31, 32).

$$a_{ij} = -\tau \frac{e}{\hbar} \int_{BZ} \frac{d\mathbf{k}}{(2\pi)^3} \sum_n \widetilde{M}_{n\mathbf{k},i} \, v_{n\mathbf{k},j} \frac{df(E_{n\mathbf{k}})}{dE_{n\mathbf{k}}},$$

$$\widetilde{M}_{n\mathbf{k},i} = S_{n\mathbf{k},i} + m_{n\mathbf{k},i}, \tag{3}$$

where $i, j, k = x, y, z$ and $\tau$ is the relaxation time. Particularly, $\widetilde{M}_{n\mathbf{k},i}$ is the magnetic moment of Bloch electrons for the band n at point **k** consisting of the spin magnetic moment $S_{n\mathbf{k},i} = <n(\mathbf{k})|\frac{1}{2}g\mu_B \sigma_i|n(\mathbf{k})>$ and orbital magnetic moment $m_{n\mathbf{k},i} = \frac{-e}{2m_e} L_n^i(\mathbf{k})$. Here, $\mu_B = \frac{e\hbar}{2m_e}$ is the Bohr magneton, g denotes the Lande g-factor set to the value 2, and $L_n^i(\mathbf{k})$ is the OAM defined in Eq. (1). Owing to the symmetry properties of the OAM, the total magnetic moment in topological chiral semimetals is zero in the equilibrium state because of cancellations between the contributions from $\mathbf{k}$ and $-\mathbf{k}$. However, an electrical current can induce an imbalance between the populations at $\mathbf{k}$ and $-\mathbf{k}$, thus generating non-zero magnetic moment and a non-zero ME susceptibility $a_{ij}$.

Constrained by the crystal symmetry in topological chiral semimetals, only three ME susceptibility tensor exists and they are equal as $a_{xx} = a_{yy} = a_{zz} = a_0$ (33). Particularly, $a_0$ is reversed by the inversion operation, giving $a_0^A \rightarrow -a_0^B$ for the opposite enantiomers. We calculated the ME coefficient, $a_0$, based on the first-principles band structure. It is noted that the relaxation time assumption in Eq. (3) limits the prediction of $a_0$. Experimentally, the observed relaxation time for topological chiral semimetals is approximately 0.1~4 ps (13, 34–38), which depends on the crystal quality, temperature, and carrier concentration. To study the effect of intrinsic band structure on the ME response of materials, the relaxation time is assumed to be 1 ps. As shown in Figure 5a and 5b, these topological chiral semimetals exhibit a large ME response. Here we assume $E_x = 10^5\ Vm^{-1}$, and the resulting total magnetization $M_0 = a_0 E_x$ can reach the values of 0.063, 0.102, 0.077, 0.029 and -0.043 $\mu_B/\text{nm}^3$ for CoSi, RhSi, PdGa, PtAl, and PtGa, respectively. The current-induced magnetization here is generally ten times larger than the reported spin magnetization in the strong Rashba systems of Au (111) and Bi/Ag (111), the surface of topological insulator α-Sn (001) surface (39–41), and comparable to the orbital magnetization in strained twisted bilayer graphene (0.20 $\mu_B/\text{nm}^2$) obtained with a large relaxation time of 10 ps,(42) as listed in Table S3. Notably, the $M_0$ varies considerably when the chemical potential is shifted. As shown in Figure 5a, the $M_0$ of CoSi, RhSi, PdGa, and PtAl increases monotonically as the chemical potential is lowered, reaching its maximum value of 0.097, 0.170, 0.353, and 0.354 $\mu_B/\text{nm}^3$ at -0.198, -0.309, -0.723, and -0.663 eV, respectively. Thus, the hole doping in these compounds is an effective way to achieve an enhanced ME response. For PtGa, it exhibits a negative $M_0$ close to the Fermi energy. The magnitude of the $M_0$ firstly increases as the chemical potential is lowered to about -0.06 eV. Further lowering of chemical potential results in a decrease of $M_0$ to 0 $\mu_B/\text{nm}^3$ at -0.186 eV, where the $M_0$ reverses sign. Then, the $M_0$ becomes positive and increases monotonically which eventually reaches the maximum value of 0.281 $\mu_B/\text{nm}^3$ at -0.624 eV. The Fermi-energy-dependent plot in Figure 5a provides a general route to optimizing the ME response in these materials.

Different ME magnitudes in five materials can be understood by $M_0$-resolved band structures as shown in Figure S5. For all these compounds, large contributions to $M_0$ were found along the $\Gamma - X$, $\Gamma - M$, and $\Gamma - R$ high symmetry lines. At the charge neutral point, the hole pockets near Γ and electron pockets near R of RhSi contribute a larger $M_0$ than CoSi and therefore RhSi shows a stronger ME response. For PtAl, PtGa, and PdGa, because the electron bands at $\Gamma$ and hole bands along the $\Gamma - \mathrm{M}$ carry negative $M_0$, PtAl, PtGa, and PdGa exhibit smaller $M_0$ than RhSi, and especially, PtGa displays a negative value.

Being directly linked to the chirality-dependent OAM texture in topological chiral semimetals, the magnetization reverses its sign for opposite enantiomers: $M_0^A = -M_0^B$, as further confirmed by the numerical calculations shown in Figure 5b. $M_0^A$ was further analyzed and its orbital ($M_0^{A,O}$) and spin ($M_0^{A,S}$) contributions were calculated using Eq. (3). As shown in Figure 5b, the orbital magnetization dominates the $M_0^A$, whereas the spin magnetization is almost negligible attributing to the antiparallel spin of spin-split bands which produces a small spin magnetization owing to the cancellations between the contributions from spin-split FSs (Figure S6 and Figure S7). The observed phenomenon is similar to the Rashba-Edelstein effect found in the noncentrosymmetric antiferromagnets (43).

To date, studies on the bulk ME effect in chiral materials are rather limited (8, 44, 45). The electric current-induced bulk magnetization was recently observed in chiral crystals of tellurium (46) and $\mathrm{CrNb_3S_6}$ (45) by nuclear magnetic resonance (NMR) and superconducting quantum interference device (SQUID), respectively. The physical origin of the observed phenomenon remains elusive, which was usually explained on the basis of the current-induced spin magnetization arising from the spin imbalance in the spin-split bands. Theoretically, we note that the ME response of some Kramers Weyl semimetals (47, 48) and chiral crystals of $\mathrm{TaSi_2}$ (49) was investigated and the mechanism based on spin/orbital texture was proposed. Our results reveal a dominating role of the exotic OAM texture of Bloch electrons for the ME effect, which is independent of the SOC strength. Furthermore, the evolution of the ME effect in different materials and the related mechanism was discussed. We found that among all five topological chiral semimetals, RhSi and PdGa are the most promising candidates for detecting the large ME response. We expect it can be experimentally measured using NMR (46), SQUID (45, 50), and Kerr spectroscopy (51). Particularly, we found that this ME response is enantiomer-dependent, manifesting as the opposite sign for opposite enantiomers, which paves a way for enantiomer recognition. Based on symmetry and numerical analyses, we propose a schematic to measure the current-induced magnetization in topological chiral semimetals, as illustrated in Figure 5c and 5d. By injecting the same electric current along the $x$-direction in PdGa enantiomers A and B, positive and negative magnetization in the $x$-direction can be observed for respective enantiomers.

In addition to the OAM-induced OHE and ME effect, the OAM can be further detected by the magnetoresistance (MR) measurement, similar to the case in the chiral-induced spin selectivity (CISS) in chiral molecules. When contacting the chiral materials with the ferromagnetic (FM) electrodes, OAM can generate the MR effect. In principle, it requires SOC to couple the spin in the FM electrode with the OAM in chiral crystals (9). Therefore, we predict that the induced MR is proportional to the SOC of the materials. MR can also be measured by applying an external magnetic field parallel to the current, which corresponds to the electric magnetochiral anisotropy (52). In this case, OAM directly interacts with the magnetic field (53). Thus, we predict that MR is insensitive to SOC here. If both MR measurements can be conducted, they will be helpful to resolve the roles of OAM and SOC in chirality-driven magneto-transport. Furthermore, angle-resolved photoemission spectroscopy (ARPES) with circularly polarized photons is also a promising route to detect the OAM. Recent studies have shown that circular dichroism (CD) can probe the chiral OAM structure in the surface states of topological insulator $\mathrm{Bi_2Se_3}$ (54). We expect that CD-ARPES can be used to probe the OAM structure in topological chiral semimetals. Moreover, the OAM texture plays an essential role in the CISS effect in molecular devices (9, 55) and anomalous circularly polarized light emission (10).

In summary, we perform both theoretical and first-principle analyses to study the OAM texture in several typical topological chiral semimetals. We find that the OAM texture displays significant chirality dependence that manifests in a sign reversal in the momentum space for opposite enantiomers. Near multifold chiral fermions, the OAM texture exhibits unique orbital-momentum locking with characteristics similar to that of the Berry curvature monopole. We demonstrate that such OAM texture in topological chiral semimetals leads to giant chirality-independent OHE and chirality-dependent OME effects, which are insensitive to the SOC strength. For five topological chiral semimetals CoSi, RhSi, PdGa, PtAl, and PtGa, RhSi shows the largest amplitudes of OHC as well as the current-induced orbital magnetization, which facilitates experimental measurements of the OHE and OME effect in this material. We expect that the induced magnetization can also induce higher-order responses such as the nonlinear anomalous Hall effect (56, 57), chirality-induced nonreciprocal magnetoresistance (58, 59), orbital current-spin current conversion (30), and exotic light-matter interaction (10).

**Methods**

All density functional theory (DFT) calculations were implemented in the Vienna ab initio simulation package (VASP).(22, 23) The exchange-correlation potential is described in the generalized gradient approximation (GGA), following the Perdew-Burke-Ernzerhof parametrization (PBE) scheme (60). The k-point grid was set to $8 \times 8 \times 8$, and the convergence of the total energy convergency was chosen to be $10^{-6}$ eV. The DFT calculations combined with the Full-Potential Local-Orbital (FPLO) package were then applied to produce highly symmetric atomic-orbital-like Wannier functions and the corresponding tight-binding model Hamiltonian (61).

**Data Availability**

All study data are included in the article and/or supporting information.

**Acknowledgments**

Q.Y. thanks Changjiang Yi and Hengxin Tan for helpful discussions. This work was financially supported by the European Research Council (ERC Advanced Grant No. 742068 `TOPMAT'). We also acknowledge funding from the DFG through SFB 1143 (project ID. 247310070) and the Wurzburg-Dresden Cluster of Excellence on Complexity and Topology in Quantum Matter ct.qmat (EXC2147, project ID. 390858490). M.G.V. and C.F. thank the Deutsche Forschungsgemeinschaft (DFG, German Research Foundation) for 5249 (QUAST). M.G.V. and I.R. acknowledge Spanish Ministerio de Ciencia e Innovacion (grant PID2019-109905GBC21). M.G.V. thanks partial support from European Research Council (ERC) grant agreement no.101020833. B.Y. acknowledges the financial support by the MINERVA Stiftung, the European Research Council (ERC Consolidator Grant ``NonlinearTopo'', No. 815869), and the Israel Science Foundation (ISF, No. 2932/21).

**Author Contributions:** C.F. conceived the project. B.Y. supervised the project. Q.Y. designed the research, performed all theoretical calculations, and analyzed the data. Q.Y. wrote the manuscript with contributions from B.Y., J.X., M.V., and I.R. All authors discussed the results and commented on this paper.

**References**

1. J. S. Siegel, Homochiral imperative of molecular evolution. *Chirality* **10**, 24–27 (1998).
2. C. Guo, *et al.*, Switchable chiral transport in charge-ordered kagome metal $CsV_3Sb_5$. *Nature* **611**, 461–466 (2022).
3. A. Inui, *et al.*, Chirality-Induced Spin-Polarized State of a Chiral Crystal $CrNb_3S_6$. *Phys. Rev. Lett.* **124**, 166602 (2020).
4. K. Shiota, *et al.*, Chirality-Induced Spin Polarization over Macroscopic Distances in Chiral Disilicide Crystals. *Phys. Rev. Lett.* **127**, 126602 (2021).
5. R. Naaman, Y. Paltiel, D. H. Waldeck, Chiral molecules and the electron spin. *Nat. Rev. Chem* **3**, 250–260 (2019).
6. Y. Wolf, Y. Liu, J. Xiao, N. Park, B. Yan, Unusual Spin Polarization in the Chirality-Induced Spin Selectivity. *ACS Nano* **16**, 18601–18607 (2022).
7. T. Liu, *et al.*, Linear and Nonlinear Two-Terminal Spin-Valve Effect from Chirality-Induced Spin Selectivity. *ACS Nano* **14**, 15983–15991 (2020).
8. F. Calavalle, *et al.*, Gate-tuneable and chirality-dependent charge-to-spin conversion in tellurium nanowires. *Nat. Mater.* **21**, 526–532 (2022).
9. Y. Liu, J. Xiao, J. Koo, B. Yan, Chirality-driven topological electronic structure of DNA-like materials. *Nat. Mater.* **20**, 638–644 (2021).
10. L. Wan, Y. Liu, M. J. Fuchter, B. Yan, Anomalous circularly polarized light emission in organic light-emitting diodes caused by orbital–momentum locking. *Nat. Photon.*, 1–7 (2022).
11. N. B. M. Schröter, *et al.*, Observation and control of maximal Chern numbers in a chiral topological semimetal. *Science* **369**, 179–183 (2020).
12. P. Sessi, *et al.*, Handedness-dependent quasiparticle interference in the two enantiomers of the topological chiral semimetal PdGa. *Nat. Commun.* **11**, 3507 (2020).
13. M. Yao, *et al.*, Observation of giant spin-split Fermi-arc with maximal Chern number in the chiral topological semimetal PtGa. *Nat. Commun.* **11**, 2033 (2020).
14. G. Chang, *et al.*, Unconventional Photocurrents from Surface Fermi Arcs in Topological Chiral Semimetals. *Phys. Rev. Lett.* **124**, 166404 (2020).
15. Q. Yang, *et al.*, Topological Engineering of Pt-Group-Metal-Based Chiral Crystals toward High-Efficiency Hydrogen Evolution Catalysts. *Adv. Mater.* **32**, 1908518

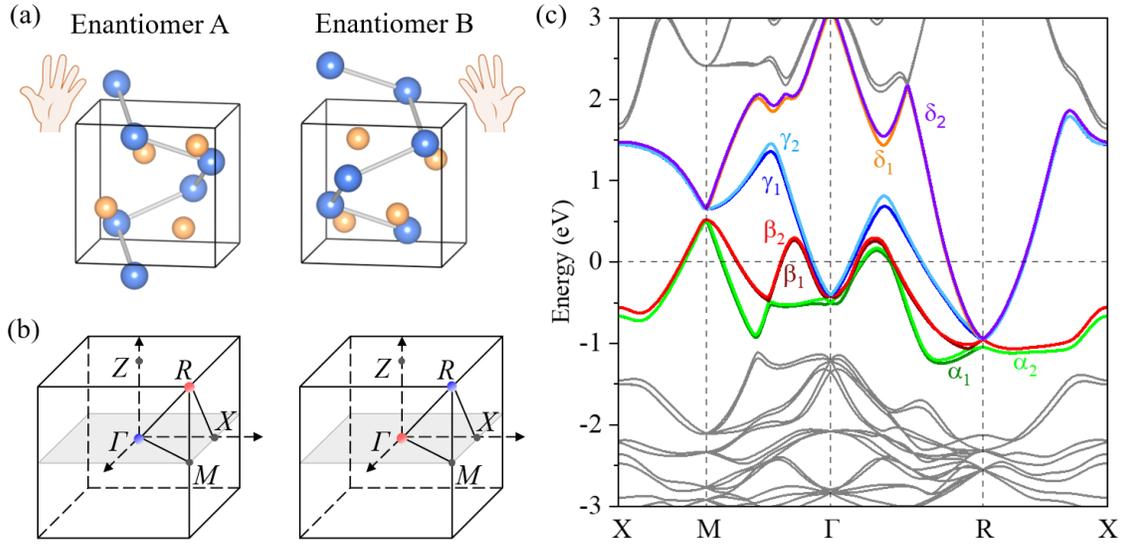

**Figure 1. Chiral crystal structure in real space and chiral multifold fermions in reciprocal space in PdGa enantiomers.** (a) Chiral crystal structures for the PdGa enantiomers (namely enantiomer A and enantiomer B). Blue and yellow atoms represent Pd and Ga atoms, respectively. Because of the $C_2$ screw rotation symmetry, the opposite chiralities can be distinguished by the helix formed by Pd atoms. (b) Cubic BZ features multifold fermions in PdGa enantiomers, which are located at the Γ/R points with Chern numbers $-4/+4$, respectively. Different enantiomers represent opposite topological charges. (c) Band structure of PdGa calculated with SOC wherein eight bands: $\alpha_1, \alpha_2, \beta_1, \beta_2, \gamma_1, \gamma_2, \delta_1,$ and $\delta_2$ intersect the Fermi energy and are marked by the colors. The subscript 1 (2) represents the down (up)-shifted spin-split bands.

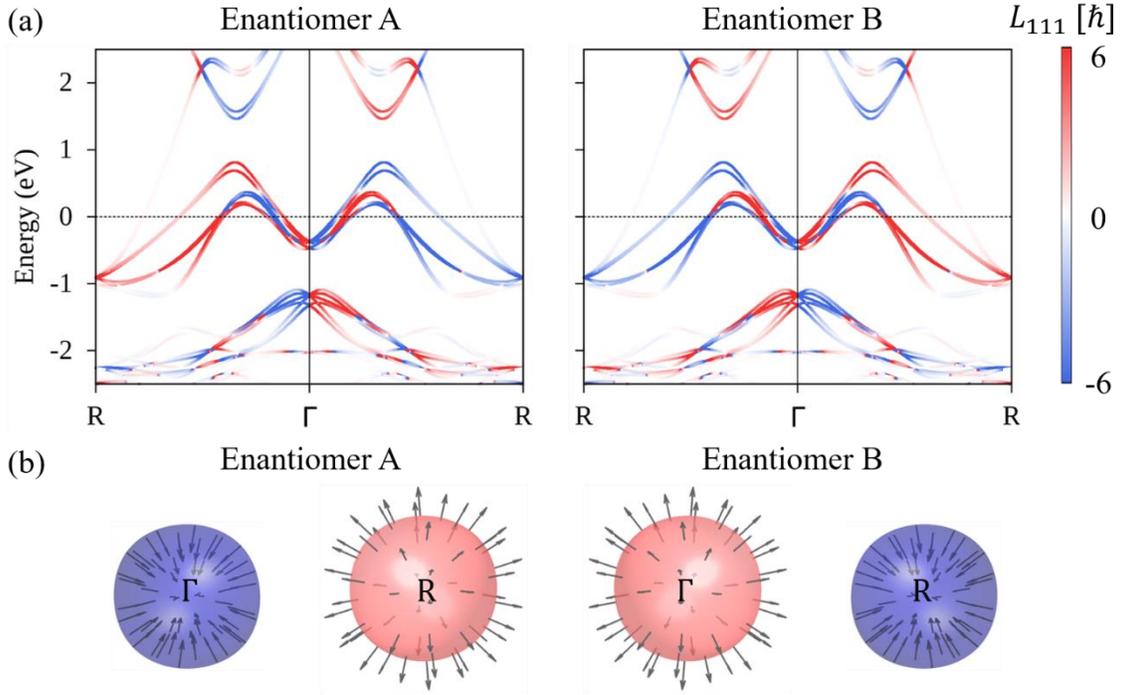

**Figure 2. OAM texture in PdGa enantiomers.** (a) Band dispersion with the OAM of $L_{111}$ for the electrons transmitted through [111] direction in enantiomers A (left) and B (right). Within one enantiomer, OAM exhibits opposite signs at $+k$ and $-k$. Furthermore, OAM is enantiomer-dependent, it reverses the sign for the opposite enantiomer. The absolute value of the OAM $|L(k)|$ is indicated in the color bar. (b) OAM texture around chiral fermions in two PdGa enantiomers. The Fermi pockets are formed by the band $\gamma_2$ centered at the Γ and R points, respectively. OAM texture exhibits monopole-like feature. The chemical potential was chosen as 30 meV above the nodes.

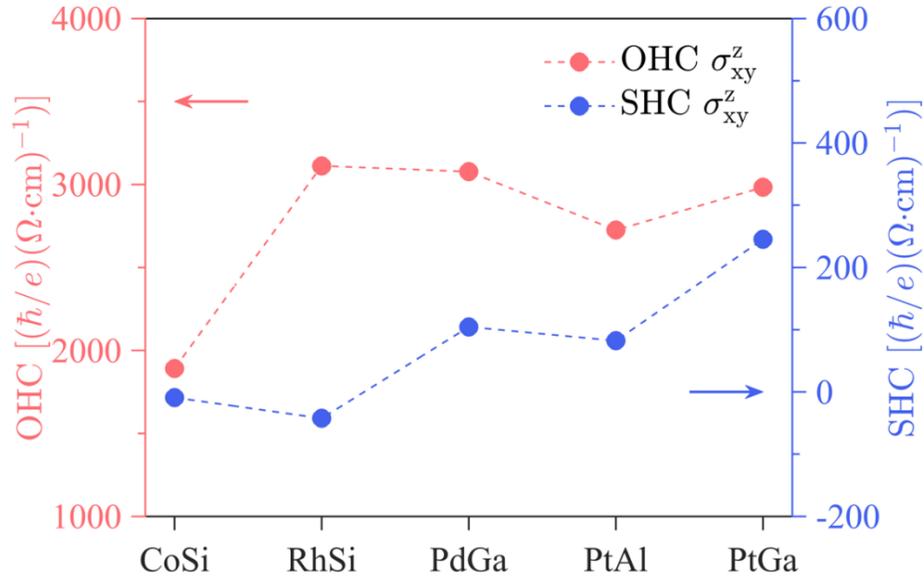

**Figure 3. Nonzero tensor element $\sigma_{xy}^{z}$ of the orbital Hall conductivity (OHC) and spin Hall conductivity (SHC) for five topological chiral semimetals.** The Fermi energy is set to the charge neutral point, which is close to the real chemical potential in materials.

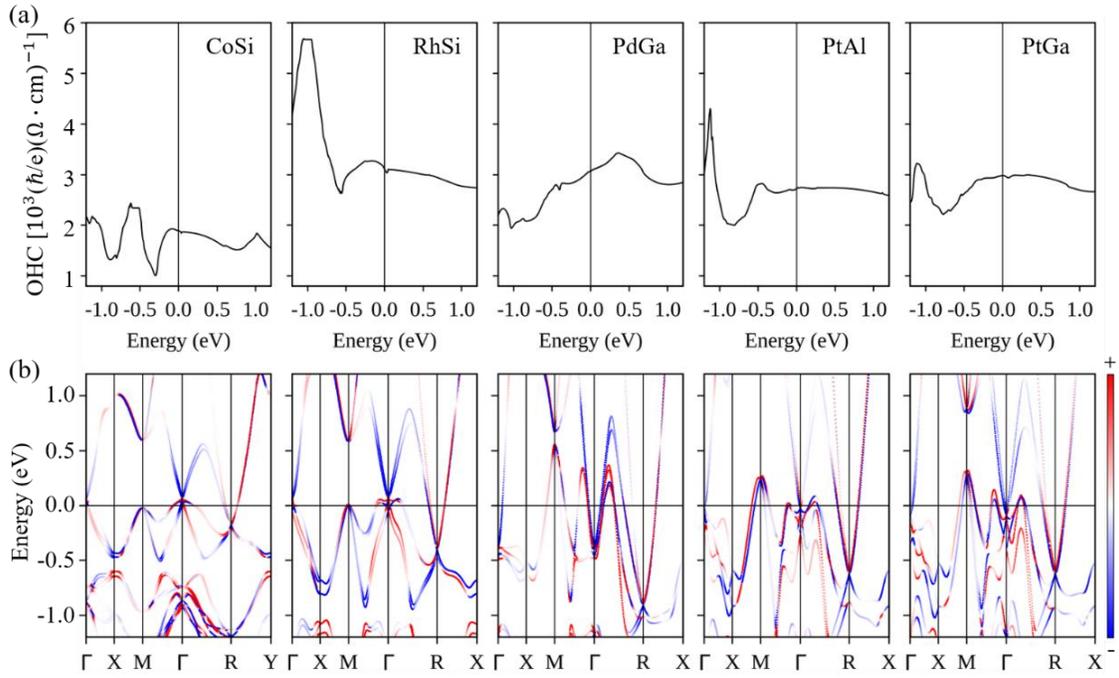

**Figure 4. Energy-dependent orbital Hall conductivity (OHC) response and the corresponding orbital Berry curvature-resolved band structure for topological chiral semimetals.** (a) OHC $\sigma_{xy}^{z}$ as a function of energy. (b) The corresponding local $\Omega_{xy}^{z}(\boldsymbol{k})$-resolved band structures along high symmetry $\boldsymbol{k}$ path.

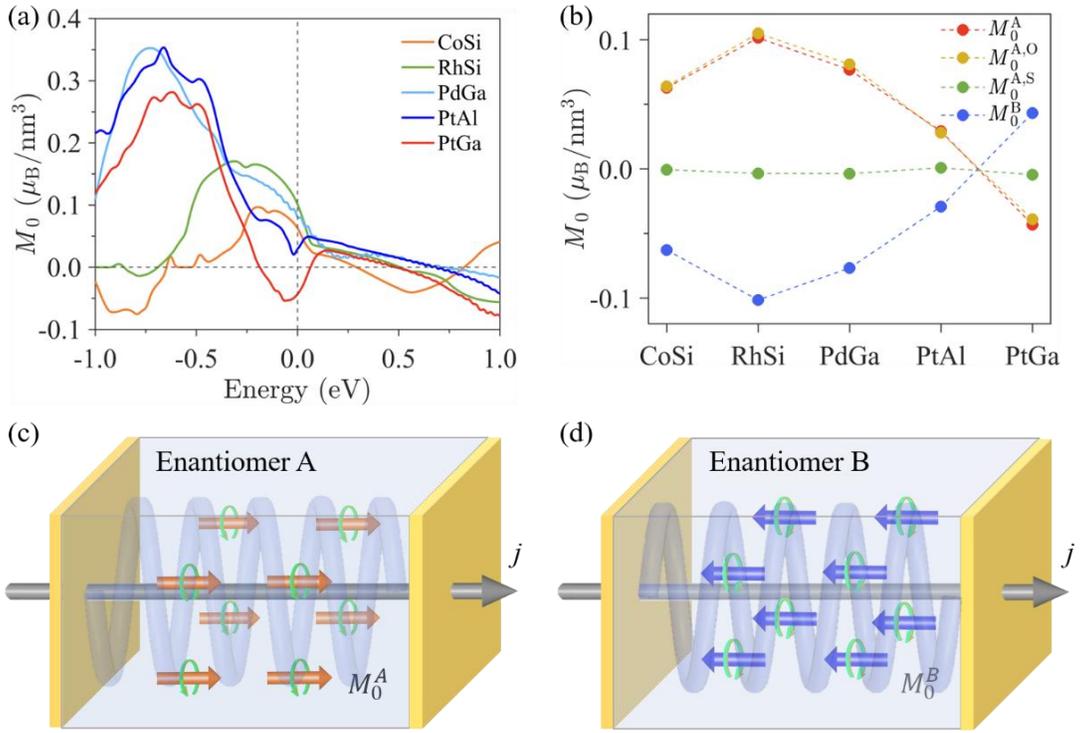

**Figure 5. The magnetoelectric (ME) response in topological chiral semimetals.** (a) Energy-dependent electric current-induced magnetization ($M_0$) in enantiomer A for all five topological multifold semimetals. (b) Electrical current-induced magnetization in enantiomers A ($M_0^A$) and B ($M_0^B$), where $M_x^A = M_y^A = M_z^A = M_0^A$ and $M_0^A = a_0^A \cdot E_x$. The corresponding orbital ($M_0^{A,O}$) and spin ($M_0^{A,S}$) contributions in enantiomer A are indicated. The total $M_0^A$ ($M_0^A = M_0^{A,O} + M_0^{A,S}$) is mainly dominated by the orbital part ($M_0^{A,O}$). The electric field is $E_x = 10^5 \, Vm^{-1}$. The Fermi energy is set to the charge neutral point. (c) and (d) are the schematics of the experimental setups to measure the chirality-dependent ME response in opposite enantiomers.